\documentclass{article}
\usepackage[utf8]{inputenc}
\usepackage{graphicx}
\usepackage{subfig}
\usepackage{float}
\usepackage{nccmath}
\usepackage{amsmath,amssymb}
\usepackage{xcolor}

\title{Phase Transitions and the Theory of Early Warning Indicators for Critical Transitions}
\author{George I. Hagstrom and Simon A. Levin\\
Department of Ecology and Evolutionary Biology\\
Princeton University}

\date{}

\begin{document}

\maketitle
\begin{abstract}
    Critical transitions, or large changes in the state of a system after a small change in the system's external conditions or parameters, commonly occur in a wide
    variety of disciplines, from the biological and social sciences to physics. Statistical physics first confronted the problem of emergent phenomena such as critical transitions in the 1800s and 1900s, culminating in the theory of phase transitions. However, although phase transitions show a strong resemblance to critical
    transitions, the theoretical connections between the two sets of phenomena are tenuous at best, and it would be advantageous to make them more concrete in order
    to take advantage of the theoretical methods developed by physicists to study phase transitions. Here we attempt to explicitly connect the theory of critical 
    transitions to phase transitions in physics. We initially find something paradoxical, that many critical transitions closely resemble first-order phase transitions,
    but that many of the early warning indicators developed to anticipate critical transitions, such as critical slowing down or increasing spatial correlations,
    occur instead in second-order phase transitions. We attempt to reconcile these disparities by making the connection with other phenomena associated with first-order
    phase transitions, such as spinodal instabilities and metastable states.
    
\end{abstract}

\section{Introduction}

Revolutions and economic collapses are some of the most dramatic and impactful historical events. They can occur with breathtaking speed 
such as the fall of Socialist governments in Eastern Europe in 1989\cite{kuran1991east} or the Black Monday stock market crash\cite{sornette2017stock}, and they often defy the expectations of
both the general public and experts, who did not foresee such sudden changes\cite{kuran1989sparks}. Although exogenous shocks can play a role in triggering large-scale social or 
economic collapse, in many cases no such shock exists and internal dynamics instead play a dominant role\cite{glance1993outbreak}. Understanding how collective behavior manifests in 
regime shifts and identifying precursors to such shifts remains an elusive challenge in social science, economics, and complex systems.

The term \emph{critical transition} describes a scenario where a small-scale shift in the forces driving a complex system leads to a large-scale change in the state
of that system\cite{scheffer2020critical}. Critical transitions commonly arise in complex systems across a number of different fields, from the social sciences and economics to ecology and
the geosciences. A simple dynamical system involving a ball rolling around a double-well potential (a landscape with two "valleys" which correspond to the wells) 
provides a simple illustration of critical transitions. A ball that is trapped in one of the valleys will remain there so long as any random perturbations do not
have sufficient magnitude to push the ball above the peak between the two wells. However, if the shape of the landscape slowly changes so that the height
difference between the well containing the ball and the peak disappears, the ball will suddenly escape from its well and roll to the other well. The resulting
change in the system state is very large compared to the infinitesimal shift in the shape of the landscape preceding it. Furthermore, if the changes to the system
reverse, so that its landscape returns to the original double-well shape, the ball will remain trapped in the second well. This hysteresis is a hallmark of many 
critical
transitions, and has profound implications: once triggered, a transition may be difficult or even impossible to reverse.

Critical transitions are an example of \emph{emergent phenomena}, which arise in systems consisting of large numbers of small-scale components (people, 
biological organisms, firms, etc) whose interactions give rise to effects occurring on a larger, macroscopic scale, which have no direct explanation
in terms of their small-scale components\cite{levin1998ecosystems}. 
    Physicists first developed a theory of emergence in the 1800s, to understand how macroscopic, thermodynamic laws arise from the microscopic interactions of matter.
    This remarkable theory, known as statistical mechanics, shows that arrangements of extremely large numbers of atoms or molecules often behave much more simply than arrangements of just a few, and that often most of the details of the intermolecular interactions disappear after the system has been coarse-grained. Statistical
    mechanics thus provides a prototype theory for the study of emergent phenomena in disciplines such as ecology or the social sciences, and statistical physics has
    found many such fruitful applications, spawning subfields such as econophysics and sociophysics\cite{Durlauf10582}. 
 
    Phase transitions in statistical physics, such as the boiling and freezing of water, the development of superconductivity or superfluidity at low temperatures, or
    the appearance of ferromagnetism in certain metals below their Curie points, are emergent phenomena that each fit the definition of a critical transition. First-order phase transitions in particular exhibit the same abrupt shift in state and subsequent irreversibility and hysteresis as classical critical transitions caused by a saddle-node bifurcation. The theory of phase transitions should thus provide insight into critical transitions.

The magnitude and irreversibility of critical transitions has motivated attempts to identify transition precursors, quantities that indicate an impending transition
when they cross a threshold\cite{scheffer2009early}. In the example with the ball on the landscape, as the well containing the ball merges with the peak separating the ball from the other valley, the curve of the landscape surrounding the ball flattens. This flattening reduces the magnitude of the restoring forces that confine the ball
to its equilibrium state, with consequences for the movement of the ball under the natural stochastic perturbations present in the system. time series measurements
of the location of the ball show \emph{critical slowing down},
a phenomenon in which the relaxation time of the ball to its 
equilibrium state diverges approaching the transition. The 
relaxation time depends on the convexity of the landscape in 
the neighborhood of the equilibrium state, and as the landscape
becomes flatter approaching the transition the strength of the
restoring force vanishes and the return time of the ball to 
equilibrium becomes infinite.

In addition to critical slowing down, the approach to the critical
transition leads to
increased variance, auto-correlation, and skewness, and increases in these quantities in a time series may thus warn of an 
impending critical transition, as has been proposed in multiple studies. These time series based early warning indicators have a rigorous mathematical basis in the
theory of bifurcations and stochastic dynamical systems, where the \emph{saddle-node} or equivalently \emph{cusp} bifurcation provides a simple normal form that
illustrates many of the key phenomena that happen in critical-transitions\cite{kuehn2011mathematical}. 

Despite the simplicity of early warning indicators as tools for detecting saddle-node bifurcations in one-dimensional dynamical systems, their application to real
world systems has yet to live up to their initial promise\cite{boettiger2012quantifying}. The reasons for this include high false-positive rates due to the \emph{prosecutor's fallacy}\cite{boettiger2012early}, a bias 
that arises from developing summary statistics based thresholds in systems already known to exhibit bifurcations, difficulties with statistical estimation, and 
the inconvenient fact that real systems may exhibit much more complexity than indicated by a simple one-dimensional bifurcation. Indeed, although using a 
one-dimensional time series or dynamical system to represent a much more complex system enables the use of early warning systems for any system where we can make
crude measurements, using a framework that embraces the high-dimensional complexity of these systems may lead to more robust early warning signals and also leverage
the growth in data gathering and analysis capabilities that has occurred in recent years.

Statistical physics, and phase transitions in particular, seem like they could provide a framework to extend early warning indicators beyond those derived from 
one-dimensional dynamical systems. However, an attempt to synthesize phase transitions with early warning indicators immediately encounters an inconsistency: critical slowing down,
increasing variance, auto-correlation, and spatial correlations all have strong analogs in the theory of second-order phase transitions (also called \emph{critical phenomena} in physics), but
other aspects of the phenomenology of these second-order phase transitions do not match what happens in a saddle-node bifurcation. 
For example, second-order phase transitions do not
exhibit irreversibility, hysteresis, or generally a large-scale jump in the state of the system. First-order phase transitions, as we alluded to earlier, do exhibit 
these characteristcs: at a first-order phase transition the state of the system jumps discontinuously, and reversing the parameters that trigger the transition do not 
immediately lead to a reversal of the transition. However, in the formalism of statistical physics, first-order phase transitions have no precursors. 

In this manuscript we have two goals: (1) propose a resolution to the inconsistency between phase transitions and critical transitions and (2) propose
early warning indicators inspired directly from the theory of phase-transitions that could portend critical transitions in economic and social systems 
that involve collective interactions. We will accomplish this by shifting to a non-equilibrium framing of phase-transitions, one which bears a much closer
resemblance to the way that critical transitions occur in the real world. In making this shift, we will see that the lack of precursors of first-order phase
transitions comes from how first-order phase transitions occur in equilibrium statistical mechanics, and that a related concept called a \emph{spinodal instability}
\cite{debenedetti1996metastable}
properly bridges between critical transitions and first-order phase transitions, applying in the non-equilibrium situations that interest us. Spinodal 
instabilities share many characteristics of second-order phase transitions and critical phenomena, including precursor effects, and thus allow for the development
of early warning indicators. This critical behavior has rarely been observed in physical systems and so has remained understudied, but its generalization to
out-of-equilibrium, athermal, or complex systems appears much more commonly\cite{abaimov2015statistical} and thus we argue that the spinodal instability is an ideal toy model or normal
form for some critical transitions that takes into consideration the full complexity of those systems. 

\section{Equilibrium Phase Transitions and Precursors}

    Statistical mechanics shows how macroscopic behavior emerges in physical systems from microscopic laws when there are large numbers of interacting molecules or
    atoms. Statistical mechanics involves a transition of viewpoints, starting with Newton's Laws of motion (or whatever is analogous for the given system), 
    but abandoning the goal of tracking the state and trajectory of every single particle in favor of a probabilistic description. The Gibbs-Boltzmann 
    distribution lies at the heart of this description: statistical mechanics postulates the probability of a given configuration of particles exchanging heat
    or other thermodynamic quantities with an environment.
    The macroscopic laws governing the system, which fundamentally relate the moments of the Gibbs-Boltzmann distribution to the parameters of the system and 
    to each-other, emerge in the thermodynamic limit, which involves allowing the system size to go to infinity. In this limit, the central-limit theorem leads to the
    suppression of fluctuations in macroscopic quantities.
  
    Before we take the thermodynamic limit, moments of the Gibbs-Boltzmann distribution are analytic functions of the temperature and other system parameters, 
    due to the analyticity of the probability distribution. Thus finite systems only have smooth changes with
    external parameters in equilibrium statistical mechanics. However, this analyticity does not always survive taking the thermodynamic 
    limit. At places where it loses smoothness, macroscopic thermodynamic variables can exhibit discontinuities or other types of non-smooth behavior. These 
    manifest as sudden changes in the macroscopic properties of the system caused by infinitesimal changes in the parameter values, and are called \textit{phase 
    transitions}. 
   
    The simplest way to illustrate phase transitions and their relationship to critical transitions makes use of the approximate Ginzburg-Landau theory\cite{ginzburg2009theory}.
    Ginzburg-Landau theory is based on the \emph{free energy}, which can be derived using the Gibbs-Boltzmann distribution and describes the thermodynamics 
    equilibria of the system (on a macroscopic scale). Ginzburg-Landau formulates the free energy of a physical system as a function of an abstract order parameter, which we imagine to be the magnetization $m$,
    an external field $h$, and the relative temperature $T$. The Ginzburg-Landau free energy takes the following form:
    
    \begin{equation}
    F(m,T) = N\mu\left(-hm+aTm^2+bm^4\right).
    \end{equation}
    
\begin{figure}
        \centering
        \includegraphics[width=2.5in]{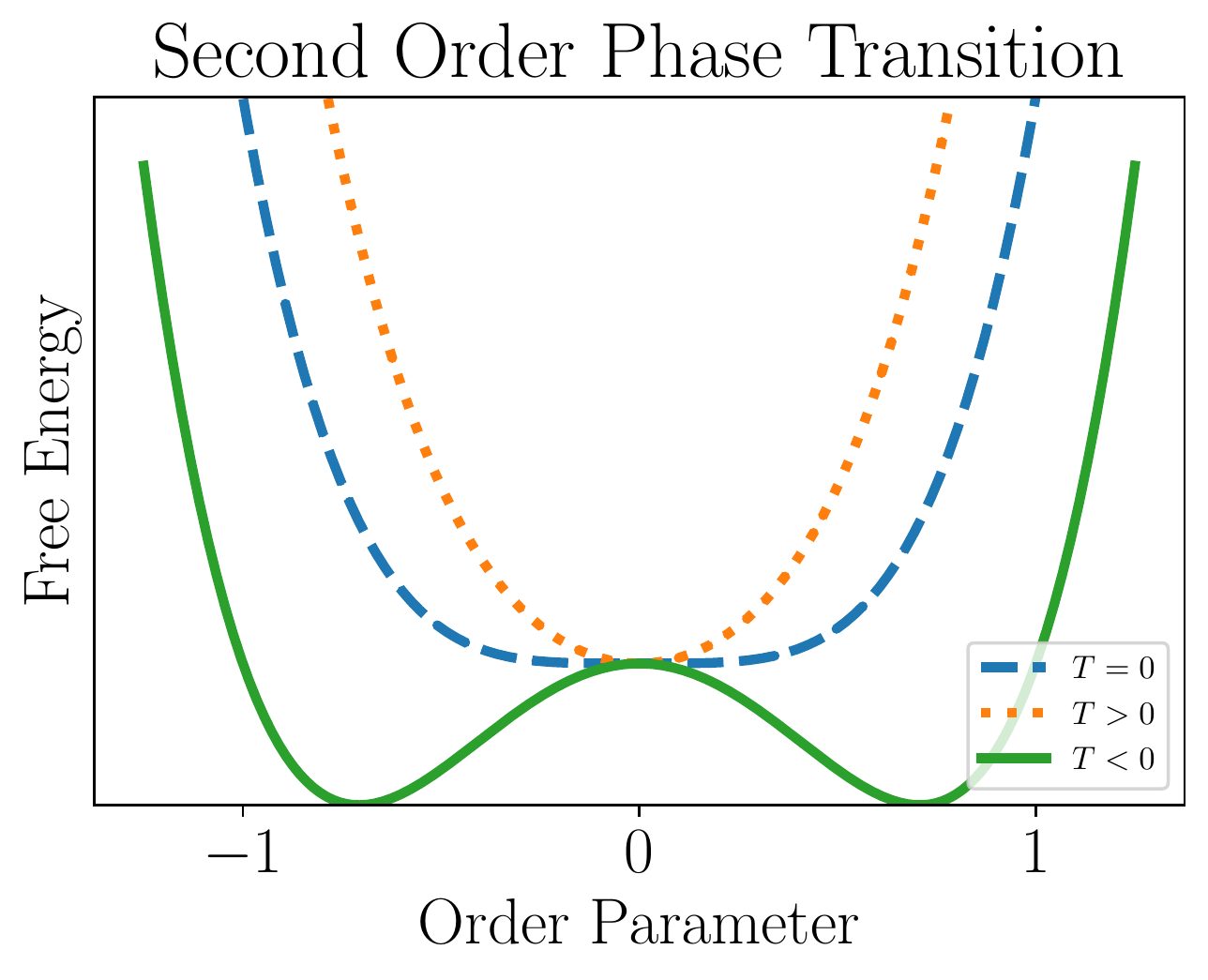}
        \includegraphics[width=2.5in]{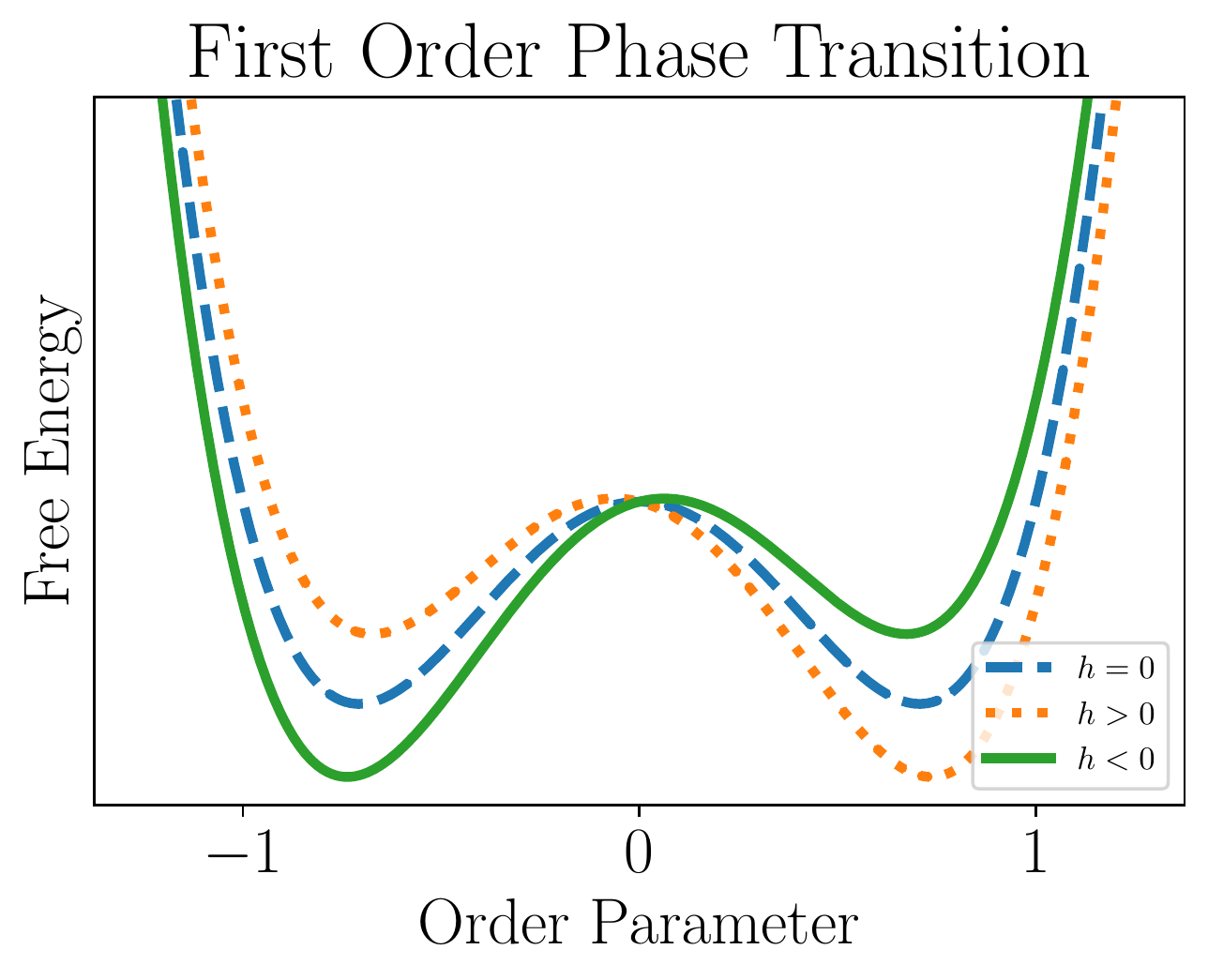}
        \caption{Phase transitions using the Ginzburg-Landau normal form. Figure (a) shows a second-order phase transition triggered by varying $T$ at zero 
        external field $h$. The free-energy undergoes a pitchfork bifurcation from a state with $m=0$ to a state with spontaneous, non-zero magnetization. At 
        the critical temperature $T=0$ the spatial and temporal correlations, the relaxation time, and the susceptibility to temperature changes and the external
        field all diverge. Figure (b) shows a first-order phase transition, driven by changing external field $h$. The transition occurs when the free energy of the 
        two global minima equal each-other. Unlike during the second-order transition, nothing special happens to the local shape of the free-energy at the 
        transition point, and thus no critical phenomena occur.}
        \label{fig:FreeEnergy}
    \end{figure}

    Here $N$ stands for the system size (number of spins in a magnetic system), $a$ and $b$ are constants, and $\mu$ is another constant, in some
    applications the magnetic
    moment of a single spin. According to equilibrium thermodynamics, the system state occupies the minimum of this free energy function. Figure \ref{fig:FreeEnergy}
    shows a plot of the free-energy for different values of $T$ at $h=0$. When $T>0$ the Free Energy has a single global minimum at $m=0$. If $T$ decreases and crosses
    through zero a pitchfork bifurcation occurs creating two new minima and converting the original minima into a local maxima, so that the full set of solutions are $m=0$ and $m=\pm \sqrt{-\frac{aT}{4b}} $. The two
    new minima have lower values of the free-energy than the unmagnetized state at $m=0$, and so the system undergoes a phase-transition at $T=0$ to either the
    positive or negative magnetization state. The development of magnetization is an example of spontaneous symmetry breaking- the Free Energy is symmetric under
    reversal of all the spins, yet the state with spontaneous magnetization must pick a sign of the magnetization. The point $T=0$ and $m=0$ is called a critical point,
    and it is characterized by a number of notable features, namely power-law scalings of system features approaching the transition.
    
    First consider fluctuations of the magnetization in a Langevin framework. We write down the time-dependent Ginzburg-Landau equation:
    \begin{equation}
    dm = - dt \frac{\partial F}{\partial m} + dW,
    \end{equation}
    where $dW$ is noise. The restoring force that maintains the magnetization at a minimum of $F$ depends on the curvature of $F$. Approaching the phase transition
    the characteristic time for return to the minimum scales as $t_{eq} \sim (|T|)^{-1}$. At the transition itself the relaxation time diverges, a phenomenon known
    as critical slowing down. Also interesting are the susceptibility of the system to changes in either $T$ or the external field $h$, both of which exhibit
    power-law divergences at the critical point with the field following the following scalings: $m\sim T^{1/2}$ and $m\sim h^{1/3}$. One can consider
    the spatial variation in the system and study the correlation length as a function of $T$, deriving in the mean field context $\zeta\sim |T|^{-1/2}$, which shows
    that approaching the critical point the system becomes scale free and spatial fluctuations occur on all length scales. 
    
    These divergences and power-law scalings bear a striking resemblance to traditional early warning indicators for a critical transition, as critical slowing down,
    rising variance and auto-correlations, and even diverging spatial correlations have all been proposed as early warning indicators. However, second-order phase
    transitions differ significantly in other important aspects from critical transitions. In particular, the order parameter undergoes a continuous, rather than
    abrupt, change at the phase transition and the transition reverses if the parameters driving the system are reversed. This suggests two things: (1) traditional
    early warning indicators require careful use and may be produced by very different phenomena and (2) we still must reconcile first-order phase transitions
    with traditional early warning indicators. We will attempt the latter in the next section on spinodal instabilities.

    At least theoretically, this intuitive expectation is correct. We call the points where the metastable state loses it stability spinodal points, and within the 
    context of mean field theory, spinodal points are characterized by much of the same phenomenology as critical points: they experience critical slowing down, they
    develop infinite susceptibility, and they develop power law spatial and temporal correlations. On this basis, we propose the spinodal instability as a potential candidate for connecting the theory of first-order phase transitions to critical transitions.

\section{First-order Phase Transitions and Spinodal Instabilities}

Consider again the Ginzburg-Landau free energy introduced in the previous section, but now assume that $T<0$ and that $h\neq 0$. Figure \ref{fig:FreeEnergy}B
shows the free energy for several different values of the magnetic field $h$. For the values shown, the free energy has two minima, one with positive and the other
negative magnetization. The magnitude of the applied field determines which is the global minimum, and thus the state at equilibrium. If the field varies from negative
to positive, then there is a first-order phase transition where the system switches from a negative magnetization to a positive one. The magnetization jumps 
discontinuously at the transition. However, no special dynamical behavior occurs at the first-order phase transition point; relaxation times, correlation lengths,
and susceptibilities remain finite. There is no sign of the impending phase transition because the shape of the local minima do not change at the
phase transition point. 

The kinetics of first-order phase transitions depend on the particular details of the system, and the transition requires time to occur.
This time depends on the height of the free energy barrier and the temperature of the system, which determines the rate of fluctuations large enough to 
push the system over the barrier into the global free energy minimum. Systems can remain trapped in these metastable states for significant, practically infinite
periods of time should the potential barrier be high enough. Diamonds provide the most famous example of such a metastable state- the graphite phase has a lower
free energy at standard tempearture and pressure but the activation energy barrier prevents the occurrence of the phase transition. 

\begin{figure}
    \centering
    \includegraphics[width=2.5in]{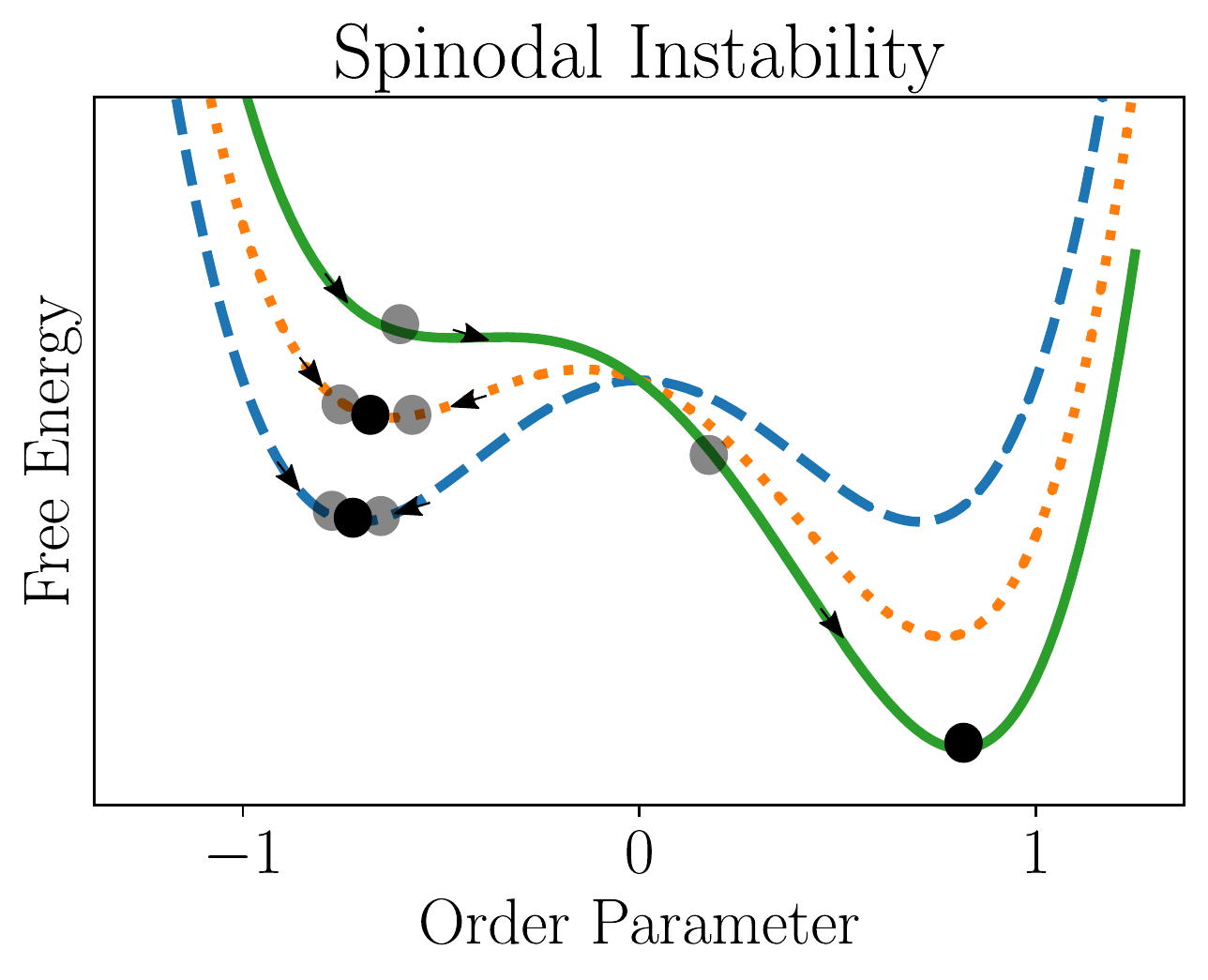}
    \caption{Spinodal Instability triggered by changing external field. As the field increases, the system undergoes a first-order phase transition but the system 
    may remain trapped in a metastable state. At a certain critical value of the field, this metastable state loses stability and the system shifts to the lowest
    free-energy state. Approaching the spinodal point, relaxation times, susceptibilities, and correlation functions may diverge, though with different
    critical exponents than during a second-order phase transition.}
    \label{fig:Spinodal}
\end{figure}

The spinodal instability provides a way for a system in a metastable state to transition to its thermodynamic equilibrium without requiring a thermal fluctuation. 
Spinodal instabilities occur when the metastable state ceases to be a local minimum of the free energy, thus causing it to lose local thermodynamic stability. 
When the system reaches the spinodal point the energy barrier disappears and the system spontaneously transitions to the equilibrium state. Figure \ref{fig:Spinodal}
illustrates the spinodal instability using the Ginzburg-Landau functional form. Here we continue increasing the external field, causing the height difference between
the two free energy minima to increase. Eventually the local minimum disappears as the local minimum and maximum collide in a saddle-node bifurcation, allowing the 
system to freely flow to the global free energy minimum.

Due to the vanishing of the second derivative of the free-energy at the spinodal point, Ginzburg-Landau theory predicts critical or critical-like behavior. 
The relaxation time diverges as $t_{eq}\sim |h-h_0|^{-1}$, the heat capacity like $C\sim |h-h0|^{-1/2}0$, the susceptibility $\xi \sim |h-h0|^{-1/2}$, and the 
correlation length as $\zeta \sim |h-h_0|^{-1/4}$. Thus, just like in the case of the critical point, a spinodal instability has a number of different precursors
that could indicate the presence of an impending critical transition.

Spinodal instabilities are the natural analogs of critical transitions based on saddle-node bifurcations 
in the context of statistical physics. They exhibit a large-scale change in system state,
hysteresis/irreversibility, and arise from the same saddle-node bifurcation within the context of time-dependent Ginzburg-Landau theory. Unlike first-order 
phase transitions, the loss of local stability leads to pronounced precursor phenomena through the critical exponents, which match our expectations derived from
early warning indicators for critical transitions. However, second-order phase transitions have qualitatively similar precursors and finding a way to distinguish 
between a spinodal instability and a critical point would be of enormous practical use.

\section{Early Warning Signals in the Fiber Bundle Model}

    Critical phenomena occupy a special role in physics, and the universality of critical phenomena has captured the imagination of scientists in a variety of
    fields, from biology to economics to the social sciences. The concept known as \emph{self-organized criticality} even posits that critical phenomena appear
    generically in complex systems, leading to the ubiquitous power law behavior observed in both natural and social systems. Spinodal criticality has drawn
    substantially less interest, and was discovered much more recently\cite{abaimov2015statistical}. This relative lack of interest likely relates to the belief that spinodal criticality
    cannot be demonstrated in real physical systems because thermal fluctuations will cause a phase transition before the system reaches the spinodal point.
    Although recent progress has been made demonstrating spinodal criticality in the superconducting Mott transition\cite{kundu2020critical}, spinodal instabilities arise regularly in 
    a wide variety of non-equilibrium systems that have applications in the social sciences and other complex systems, in particular \emph{athermal} systems 
    with \emph{quenched disorder}, and other non-equlibrium systems associated with damage phenomena. 
 
\begin{figure}
    \centering
  \centerline{  \includegraphics[width=5in]{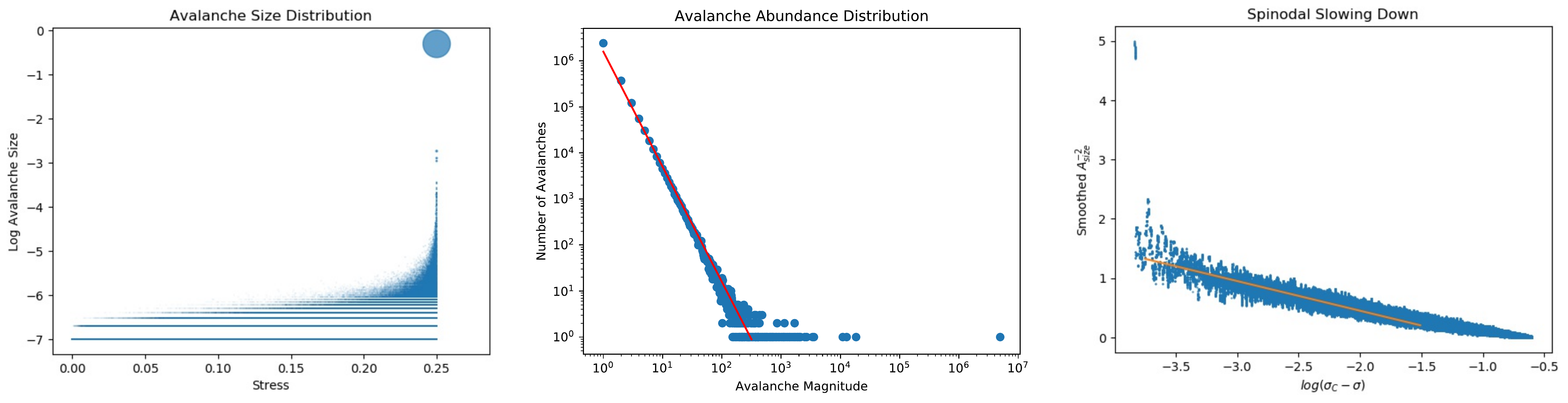}}
    \caption{Avalanche behavior and critical exponents in the Democratic Fiber Bundle Model: (left) As stress increases, avalanche size increases, until a single avalanche orders magnitude 
    greater than all preceding avalanches causes catastrophic failure of the entire bundle. (center) The avalanche magnitude-abundance distribution shows a -5/2nd power law scaling. (right) The size of the mean avalanche increases approaching the spinodal point with a critical exponent of -1/2.}
    \label{fig:FiberBundle}
\end{figure}
   
    Spinodal criticality was first used to understand and predict impending transitions in geophysics. Earthquakes exhibit behavior characteristic of both
    first and second-order phase transitions, most famously the power-law distribution of earthquake magnitudes known as Richter's law\cite{rundle2000precursory}. 
    Self-organized criticality offered a potential explanation, suggesting that the constant development of stresses in the Earth's crust drives 
    the system to the critical point continually, with stress released through periodic earthquakes. However, 
    the sudden drop in stress after an earthquake is much more characteristic of the discontinuous change at a 
    first-order phase transition. Rundle et. al. \cite{rundle2000precursory} developed a theory of earthquakes as damage phenomenon described by a 
    spinodal instability. Their theory predicted that an avalanche cascade
    characterized by critical exponents coming from the theory of spinodal instabilities precede large earthquakes, thereby acting as a means of predicting them.
    
    Athermal systems with quenched disorder also manifest avalanche cascades with such properties\cite{zapperi1997first,abaimov2015statistical}. 
    A classic example is the phenomenon of Barkhausen noise, which
    are the avalanches of spin flips that occur in a magnet under an applied field. The zero-temperature Random Field Ising Model describes Barkhausen avalanches
    and has also been generalized to apply to a wide variety of sociological and economic phenomena, such as market crashes, spontaneous adoption of new
    technologies, or even political revolutions\cite{bouchaud2013crises}. Here, however, we focus on an even simpler example of the spinodal instability, the {fiber bundle model} (FBM).

The FBM provides a simple example of an exactly soluble, athermal model exhibiting spinodal criticality and large-scale failures. Originally
invented to model the pattern of fiber failures and the eventual breakage of materials placed under increasing strain, the fiber bundle model has applications well
beyond materials science, and it has been used to study fragility in food networks and other supply chains as well as collective social phenomena such as the sudden
adoption of a behavior, for example the formation of lines before boarding gates at airports or the cessation of applause following concerts. The fiber bundle model
bears a close relationship to other important athermal systems which exhibit avalanche cascades and have been used to study social dynamics, such as the 
Random Field Ising Model. Here we select it to illustrate early warning indicators because of its relative simplicty and broad applicability.

In its materials science formulation, the fiber bundle model describes a material composed of a large number $N$ of individual fibers. The ends of the material
experience a force, causing the fibers to elongate according to Hooke's law. In the \emph{democratic} fiber bundle model (DFBM), the strain in the material distributes equally
among the fibers, so that the elongation relates simply to the number of fibers and the total strain:
\begin{equation*}
    \sigma = \frac{kx}{N}.
\end{equation*}

Here the spring constant $k$ relates the strain $\sigma$ to the elongation of the fibers $x$. Each fiber has a maximum elongation, beyond which it ruptures. The
elongation threshold $x_c$ for each fiber follows a probability distribution $p(x_c)$. Ruptured fibers do not bear any strain, and their breaking causes a 
redistribution of the load to the remaining fibers. Depending on the external strain $\sigma$, the material may come to equilibrium with some fraction of its fibers
broken, but still able to support the load, or if the load exceeds a certain critical threshold, all of the fibers will break and the material will fail 
catastrophically. 

Consider a material experiencing a quasistatic increase in external stress. As the stress increases, fibers will begin to elongate beyond their thresholds and thus
rupture. These ruptures will further increase the elongation, which may push other fibers across their thresholds and so on. The result is that as the stress slowly
increases, a series of avalanches occur which slowly damage the material. Near the critical level of stress, one giant avalanche destroys a significant fraction of the fibers in the material, causing its failure. This catastrophic avalanche represents a spinodal instability- the stress bearing state of the material is metastable,
and at the failure point the system transitions to the broken state. In the vicinity of this spinodal point, the avanlanches exhibit spinodal criticality. The 
mean avalanche magnitude diverges at the spinodal point with a critical exponent of $-\frac{1}{2}$:
\begin{equation}
N_A \sim (\sigma_c-\sigma)^{-1/2}.\label{eq:CriticalExponent}
\end{equation}

Fig. \ref{fig:Spinodal} illustrates an avalanche cascade in the DFBM with a uniform distribution of failure points. The power law scaling appears well before the
actual spinodal point, suggesting the possibility of detecting the imminent collapse as proposed by Pradhan et. al.\cite{pradhan2001precursors} (and by Rundle et. al. \cite{rundle2000precursory} in the specific context of earthquake prediction) who suggested using
the power law behavior in a variety of complexity systems which exhibit either self-organized criticality or the spinodal instability as precursors of large-scale
disruptions or critical transitions. 

Here we develop a system to predict the time of catastrophic failure in the DFBM from the time series of avalanches. Instead of using threshold or summary statistics
based approaches, which suffer from the difficulty of selecting an appropriate signficiance threshold, poor null models, low statistical power, and an inability to quantify uncertainty\cite{boettiger2012quantifying}, we instead employ a model based system that fits the avalanche time series either to a distribution which has an explicit power-law term to compute a posterior distribution of blow-up times and use information-criteria based model comparisons to compare with other possible
null models. Specifically, given a time series of avalanche sizes $N_t$, we partition the time series into a large number of discrete windows and fit the 
avalanche distribution in each interval to a Borel distribution (avalanches at fixed strain follow a Borel distribution\cite{hansen1992distribution}) using maximum-likelihood methods. We then fit two alternative models to the resulting time series, a blow-up model that includes a term proportional to 
Eq. \ref{eq:CriticalExponent}, and a null model that fits a parabolic function to the time series. The blow-up model takes the form:
\begin{equation*}
    \bar{N_{A}}(t) \sim C_0 + C_1 t + C_2 (t_c - t)^{-1/2} + \mathrm{normal}(0,\epsilon),
\end{equation*}
while the null model obeys:
\begin{equation*}
    \bar{N_{A}}(t) \sim C_0 + C_1 t + C_2 t^2 + \mathrm{normal}(0,\epsilon).
\end{equation*}
We fit the parameters of each of these models to the time series using a Bayesian approach, implementing the models in the Stan probabilistic programming language
\cite{stanmanual,carpenter2017stan}. In the case of the blow-up model, we determined from partial time series in the lead-up to the catastrophe the posterior 
probability of each catastrophe time by marginalizing over the other parameters. We used the the Widely Applicable Information Criterion (WAIC)\cite{watanabe2010asymptotic} to compare the null model to the blow up model, which we computed using leave-one-out cross validation\cite{vehtari2017practical}.
This allows for comparison of models with different numbers of parameters. When the null model has a similar statistical weight to the blow-up model, we conclude
there does not exist sufficient evidence in the time series to predict a catastrophe.

\begin{figure}
    \centering
    \centerline{\includegraphics[width=6.0in]{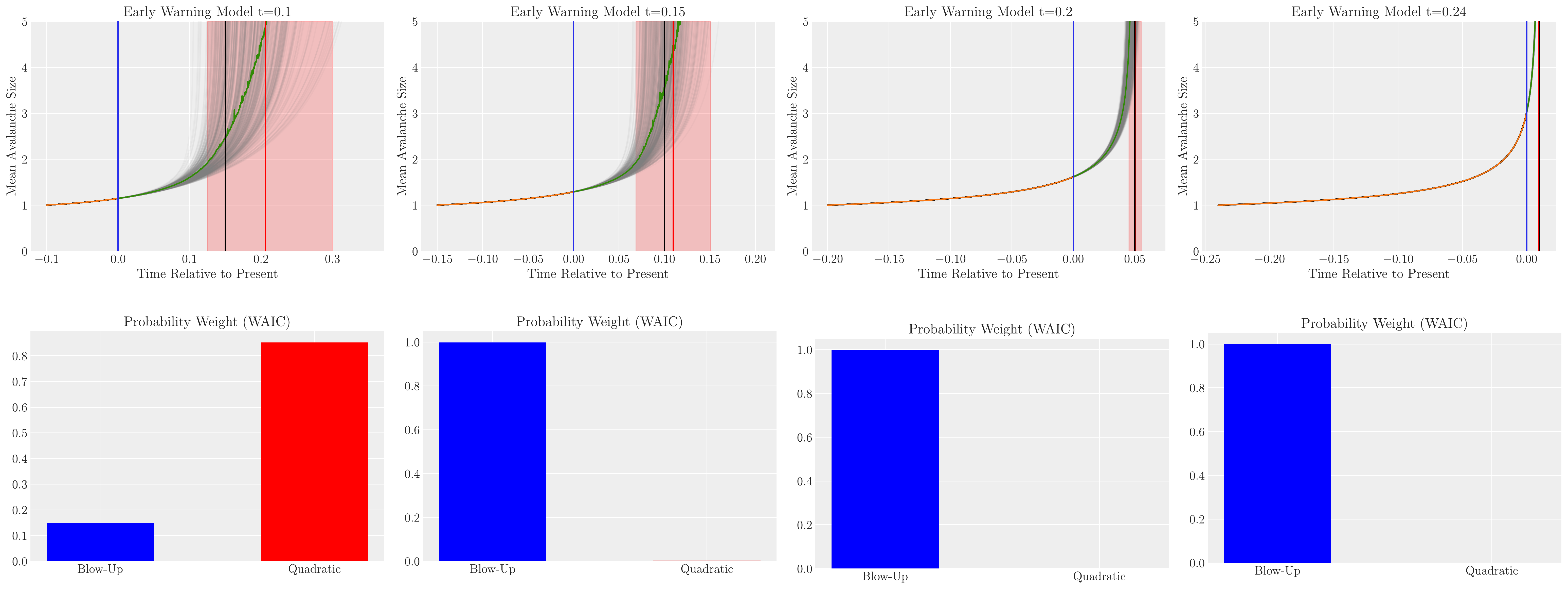}}
    \caption{Prediction of catastrophic failure times in the DFBM from time series data. Each column shows draws from the posterior distribution of predictions
    of the mean avalanche size from a time series. All the time series originate from the same numerical simulation, but moving right across the chart, each includes more data.
    Catastrophic failure occurs at $t=0.25009$, the first time series cuts off at $t=0.1$, the second time series at $t=0.15$, the third at $t=0.2$, and the
    final at $t=0.24$. The blue line marks the point in time at which the time series stops, and the trajectories to the right of the blue line are model 
    predictions drawn from the posterior distribution. The red lines are the mean prediction for when failure occurs, the red area a 95\% credible interval of the
    failure time, and the black line marks the actual model failure location. The second row shows a model comparison between the blow-up and a quadratic model.
    As the final point of the time series gets closer to failure, the predictive accuracy improves rapidly. For the first time series, the quadratic model performs
    better than the blow-up model, but at all subsequent points the blow-up model has a much greater probabilistic weight, suggesting that blow-up has statistical support.}
    \label{fig:BlowUp}
\end{figure}

Figure \ref{fig:BlowUp} shows how the posterior distribution of the failure time $t_c$ evolves as the time series advances, and how model quickly the 
WAIC can identify the correct model. Even a fair distance from the transition, the posterior begins to converge around the correct time, and all of the weight transfers
to the blow-up model. Using the model based on the correct normal form for the bifurcation, we can provide early warning of catastrophic failure with robust measures 
of statistical uncertainty.

\section{Conclusions and Prospects}

Critical transitions have always shared a close, but imprecise connection with statistical physics and phase transitions. Here we proposed making the analogy
between the two concepts more direct, associating critical transitions in complex systems with the spinodal instability of statistical physics. This interpretation
simplifies several seeming inconsistencies between the two concepts: the mismatch between the abrupt nature of critical transitions, which resemble 
first-order phase transitions, with the appearance of power laws and other precursor phenomena which normally occur only near critical points at second-order
phase transitions. Spinodal instabilities, which occur when a metastable state loses stability, causing an abrupt transition to a thermodynamically stable state,
simultaneously exhibit both characteristics of first and second-order phase transitions, though the criticality observed near spinodal points has received much less
attention in the physics literature due to difficulties observing it experimentally.

However, it has been known for some time that spinodal criticality can be easily seen in so-called athermal systems with quenched disorder, such as the 
Democratic Fiber Bundle Model and the Zero-Temperature Random Field Ising Model, models which have been widely applied outside of physics to social, economic, and 
ecological phenomena. In particular, the abrupt transitions that occur in these systems can model the large-scale collapse of complex social systems such as governments
or trade networks.

Much of the research on critical transitions centers on the idea that they may exhibit precursory phenomena that enable their prediction. One of the strongest 
potential applications of the physics of the spinodal instability is that it provides a principled means of deriving early warning indicators based on the
critical exponents near the spinodal point. Here we demonstrated the feasibility of this idea by building a model-based early warning indicator for the catastrophic
failure of the fiber bundle model under increasing stress, which predicts a probability distribution for failure times and compares with a null-model that has no 
failure. Although the early warning indicator we proposed here is relatively simple, based on scaling of the avalanche size distribution near the spinodal point, 
more complex systems will possess a variety of critical exponents which can be simultaneously studied. Universality is a powerful phenomenon which causes seemingly 
different systems to have identical critical exponents, a tool which can be leveraged to extend knowledge between collapse in different systems. 

Making a connection between critical transitions and phase transitions will help realize the tremendous promise in the theory of early warning indicators for 
critical transitions, but substantial challenges still remain. The ideas presented here relied heavily on mean field theories, which make for simpler analysis but 
can sometimes behave differently than low dimensional systems near spinodal points. Although the spinodal instability clearly arises in the special case exhibited here
(and in other cases of interest), it is possible that some of the associated critical phenomena become difficult to observe in real world systems. The power laws
that we proposed to indicate an approaching spinodal instability can also indicate the presence of a regular critical point, which would not lead to an abrupt 
transition or an irreversible catastrophe. How easy is it to distinguish between spinodal instabilities and regular criticality? Lastly, the spinodal instability requires testing outside the context of systems from physics or materials science. Will social or economic systems provide enough data to fit models of the type 
that we proposed to use? Some promising work exists suggesting that it is possible to detect critical exponents indicative of a spinodal point in low-data systems
such as the collapse of applause after a concert, and the dramatic increase in the availability of social and economic data in recent years gives us hope that we 
can find plenty of data rich systems on which to test our hypotheses. 

\section{Acknowledgements} We would like to thank Pablo Debenedetti for helpful discussions on the physics of phase transitions and spinodal points. We also acknowledge gratefully support from the Army Research Office  grant – W911NF-18-1-0325, the National Science Foundation Grants DMS 1951358, CCF1917819, and OCE1848576, the National Oceanic and Atmospheric Administration grant  NA18OAR4320123, and DARPA Young Faculty Award number N66001-17-1-4038.

\bibliography{library}
\bibliographystyle{abbrv}
\end{document}